# Identifying Helpful Context for LLM-based Vulnerability Repair: A Preliminary Study

Gábor Antal
gabor.antal@frontendart.com
antal@inf.u-szeged.hu
FrontEndART Ltd.
University of Szeged
Szeged, Hungary

Bence Bogenfürst
bogen@inf.u-szeged.hu
University of Szeged
Szeged, Hungary

Rudolf Ferenc
ferenc@inf.u-szeged.hu
University of Szeged
Szeged, Hungary

Péter Hegedűs
peter.hegedus@frontendart.com
hpeter@inf.u-szeged.hu
FrontEndART Ltd.
University of Szeged
Szeged, Hungary

## Abstract

Recent advancements in large language models (LLMs) have shown promise for automated vulnerability detection and repair in software systems. This paper investigates the performance of GPT-4o in repairing Java vulnerabilities from a widely used dataset (Vul4J), exploring how different contextual information affects automated vulnerability repair (AVR) capabilities. We compare the latest GPT-4o's performance against previous results with GPT-4 using identical prompts. We evaluated nine additional prompts crafted by us that contain various contextual information such as CWE or CVE information, and manually extracted code contexts. Each prompt was executed three times on 42 vulnerabilities, and the resulting fix candidates were validated using Vul4J's automated testing framework. Our results show that GPT-4o performed 11.9% worse on average than GPT-4 with the same prompt, but was able to fix 10.5% more distinct vulnerabilities in the three runs together. CVE information significantly improved repair rates, while the length of the task description had minimal impact. Combining CVE guidance with manually extracted code context resulted in the best performance. Using our Top-3 prompts together, GPT-4o repaired 26 (62%) vulnerabilities at least once, outperforming both the original baseline (40%) and its reproduction (45%), suggesting that ensemble prompt strategies could improve vulnerability repair in zero-shot settings.

## CCS Concepts

• **Security and privacy** → **Software and application security**; • **Software and its engineering** → **Software post-development issues**.

## Keywords

LLM, vulnerability repair, contextual information, prompt engineering, software security, automated program repair

**ACM Reference Format:**
Gábor Antal, Bence Bogenfürst, Rudolf Ferenc, and Péter Hegedűs. 2025. Identifying Helpful Context for LLM-based Vulnerability Repair: A Preliminary Study. In *Proceedings of The 29th International Conference on Evaluation and Assessment in Software Engineering (EASE 2025).* ACM, New York, NY, USA, 6 pages. https://doi.org/XXXXXXX.XXXXXXX

## 1 Introduction

The rapid advancement of large language models (LLMs) has opened new possibilities for automated vulnerability detection and repair in software systems. Recent studies [17, 19] have demonstrated that LLMs can achieve notable success in fixing real-world vulnerabilities, although with significant limitations.

A crucial aspect of automatic vulnerability repair (AVR) is the availability of relevant contextual information. In its most basic setting, the models are only aware whether a given piece of code contains a vulnerability, but they miss any detailed descriptions regarding its nature. Supplementary sources of contextual information, such as Common Weakness Enumeration (CWE) [5] labels, issue descriptions, or Common Vulnerabilities and Exposures (CVE) [4] reports may enhance the model's ability to generate correct fixes. However, such additional information is not always available in a real-world environment.

The use of retrieval-augmented generation (RAG) techniques has been explored in previous works [7, 13] for vulnerability detection and repair. However, when extensive vulnerability-fix datasets are not readily available (i.e., using RAG is limited), alternative sources of information, such as related repository documentation or library usage examples, could serve as valuable contextual sources. In many cases, the necessary information to fix a vulnerability already exists within the project's code base. However, unless explicitly guided, an LLM may not have knowledge of these existing solutions. Due to technical limitations, it is not possible for the model to process the entire project repository. The data provided (including code snippets) must be carefully composed to include only the most relevant information. Extracting such relevant information automatically based on the vulnerable code remains a challenge.

In this preliminary study, we manually collect relevant context from the repository or other sources to evaluate its impact on the repair process. Although this manual approach might introduce subjectivity, it serves as a first step towards the development of an automated tool that systematically retrieves project-specific repair knowledge. Previous works [9, 11] have already shown how increasing contextual information improves repair rates, however, these works have not investigated how different contextual elements affect the results.

In this paper, we investigate the performance of the GPT-4o large language model in repairing Java vulnerabilities. We use prompts with different kinds of contextual information. For our evaluation, we chose the Vul4J [3] dataset, a collection of reproducible real-world Java vulnerabilities. We analyze whether providing additional details, such as vulnerability descriptions from CVE reports and manually identified relevant code snippets, can improve the model's repair capabilities. Furthermore, we used the prompt of a previous work [17] and compared its results with a newer model (GPT-4o) to see whether more advanced models can achieve higher repair rates with the same prompt. In our work, we answer the following research questions:

- **$RQ_1$**: How does different contextual information affect the vulnerability repair performance of GPT-4o?
- **$RQ_2$**: How does the use of multiple different prompts contribute to the overall success rate of vulnerability repair?

Comparison of the performance of GPT-4 and GPT-4o using the same base prompt from our previous work [17] revealed that GPT-4o performed on average 11.9% worse, suggesting that the more advanced models are not guaranteed to perform better in repair tasks or require different composed prompts. However, considering the number of distinct vulnerabilities successfully repaired in at least one of the three runs, GPT-4o achieved 10.5% better results.

We conducted a study on 42 vulnerabilities in the Vul4J dataset with nine prompts that contain different kinds of contextual information. We generated a fix with each prompt three times using GPT-4o (1,134 fixes in total) and evaluated them using the unit tests included in Vul4J. Our experiments show that task description length does not affect repair ability, but CVE-related information significantly improves fixing rates. Combining CVE guidance with manually extracted code context achieved our highest repair rates (18-14-17 over three runs). Using our Top-3 prompts, GPT-4o repaired 26 (61.9%) distinct vulnerabilities at least once, outperforming both the original baseline (40.4%) and its GPT-4o reproduction (45.2%). This indicates that using multiple prompts (with various context information) is a promising way to increase LLM repair rates.

## 2 Related Work

Recent advances in Large Language Models (LLMs) have demonstrated potential for automated vulnerability detection and repair. Previous studies evaluated the zero-shot repair capabilities of LLMs on real-world vulnerabilities, showing that LLMs can produce fixes but often require iterative prompting and external validation [9, 14, 17]. Our study builds on this by evaluating GPT-4o for the repair of Java vulnerabilities using Vul4J and exploring how contextual information affects the effectiveness of the repair. Retrieval-augmented generation (RAG) approaches enhance LLMs by incorporating external knowledge [8].

Vulnerability repair requires validation mechanisms such as unit tests or fuzzing frameworks to ensure correctness. Previous works have introduced automated repair pipelines that incorporate test validation and iterative refinement [9, 21]. Benchmarks like Vul4J [3] and Big-Vul [10] provide real-world datasets to evaluate vulnerability repair.

Prompt engineering might also be crucial for improving LLM-based AVR. Chain-of-Thought (CoT) prompting has been shown to enhance LLM reasoning for various tasks [18, 20, 22]. Instruction-based prompting further refines LLM outputs, as demonstrated in [1, 6]. In our work, we investigate how different contextual prompts (CWE/CVE information, extracted code context) affect repair rates and find that CVE-related data significantly improve performance.

Different LLMs have been benchmarked for security-related tasks. GPT-4 has outperformed fine-tuned models like CodeT5 and CodeBERT on vulnerability detection and repair tasks [14, 17]. Studies comparing open and closed source models suggest that fine-tuned models can reach GPT-4 performance with sufficient domain-specific training [9]. Our work contributes to this discussion by showing that specific contextual information can influence the result of a given model.

## 3 Methodology

### 3.1 Prior Study Replication

To assess whether LLM-based vulnerability repair has improved with recent model advancements, we first replicated our previous work [17]. Specifically, we executed the same prompts using OpenAI's GPT-4o model.[1] The temperature parameter was set to 0.01 for all runs, mirroring the base study's configuration. To align with the established methodology, we generated code 3 times with each prompt. Due to reproducibility challenges, our analysis includes only 42 vulnerabilities, rather than the 46 in the base study. The authors also encountered similar issues, and not all 46 cases could be verified using unit tests. For consistency, our comparison uses the same subset of 42 projects. For some vulnerabilities that could not be executed in the original study, we managed to fix the environment, and run the unit tests. Thus, using the responses they published, we were able to run the generated code fixes in these cases.

*3.1.1 Evaluation.* For efficient evaluation, we developed an automated tool that integrates the generated code snippets into the appropriate locations within the original projects, aiming to minimize the need for manual inspection. The base study noted that certain test cases were overly strict, potentially misclassifying correct fixes as failures. Where possible, we improve the tests (VUL4J-5, VUL4J-13, VUL4J-39, VUL4J-41, and VUL4J-48) to relax such strict constraints while preserving their core validation logic. Furthermore, for four projects (VUL4J-16, VUL4J-53, VUL4J-63, and VUL4J-65), we performed manual inspections, as these cases presented challenges within the test framework or in the automated insertion process. For the remaining projects, we exclusively relied on automated testing. Manual verification was carried out only in instances where the baseline model (GPT-4) succeeded in repairing a vulnerability, but GPT-4o did not, or when a compilation error occurred. The generation was executed three times independently (without changing run parameters), to account for the variability of the response.

In summary, GPT-4o successfully repaired 10, 12, and 14 projects across the three runs, which is slightly lower than GPT-4's reported results (14, 15, and 13 successful fixes, respectively). These results suggest that GPT-4o does not bring about improvements in the AVR task by default, without explicitly leveraging its capabilities. Therefore, it is important to further explore the appropriate contextual enhancement techniques to improve results.

---
[1]The exact version was *gpt-4o-2024-08-06*.

## 3.2 Additional Context

Given the baseline results, we hypothesized adding contextual information could enhance the model's repair capabilities. Our initial approach focused on modifying the code input.

*3.2.1 Increased code visibility.* The base prompts included only the parent class header, all data members, and the vulnerable method, but omitted other class methods. This often led GPT to erroneously reimplement already existing internal helper functions, potentially degrading the quality of the generated fixes. To mitigate this, we include the function headers of all class methods referenced within the vulnerable function. However, the actual implementations of these methods were redacted with comments (`"IMPLEMENTATION REDACTED"` and `"IMPLEMENTATION IS SECURE"`). This ensured that the model would not generate unnecessary replacements but could still understand the method dependencies within the class.

*3.2.2 Restoring documentation comments.* The original prompts removed the method-level documentation comments. We opted to reintroduce them whenever available, as the original documentation often contained project-specific information that could help the model to understand the code.

*3.2.3 Refining contextual relevance.* We excluded irrelevant data members from the provided code snippets to reduce noise in the input of the model. Additionally, we explicitly included necessary import statements to prevent automatic code insertion failure, as GPT did not consistently generate missing imports in its responses. This would also help the model understand the usage of statements.

These refinements improved consistency in the model output formatting, which in turn facilitated a more reliable integration of the generated fixes into affected projects.

To further support the model's comprehension of the vulnerable code, we generated line-by-line documentation for each vulnerable method. Inspired by previous studies on LLM-generated code summaries [7, 20], we used GPT to annotate functions with explanatory comments above each major statement. Examples of generated comments can be found in our online Appendix.[2] The rationale behind this approach was that improving the model's understanding of the program's semantics would lead to better-informed and more accurate vulnerability repairs.

We found that placing the example code snippet, which demonstrates the desired comment generation style, inside the system prompt led to the most accurate and consistent inline documentation. This approach ensured that the model adhered more closely to the desired formatting and level of detail in its responses. We have built our subsequent prompts based on these insights, relying more on the system prompt.

## 3.3 Prompt Engineering

To further optimize repair performance, we experimented with different prompt engineering approaches inspired by strategies listed in the work of Pranab et al. [16]. Given our constraints, we prioritized techniques that could be utilized in a zero-shot setting [15] or those that allowed the necessary contextual information (e.g., documentation or details related to CVE) to be generated in advance. Using insights from other works [2, 12, 22], we constructed an initial comprehensive system prompt, which included detailed instructions and best practices for vulnerability repair. The corresponding user prompt retained the same core strategies, but was formulated in a more concise format to improve efficiency. We hypothesized that balancing prompt length and instructional clarity could impact the quality of model response.

To further refine our prompts, we engaged GPT itself in the prompt engineering process. Specifically, we tasked GPT [6] with reorganizing and improving our initial manual prompts. These refined prompts aimed to improve instruction clarity and guide the model toward more reliable vulnerability repair. However, our empirical testing revealed no significant differences in the result quality between the longer, more detailed prompt and shorter, more condensed variant. Given this observation, we opted for the shorter prompt in subsequent experiments, as it reduces unnecessary complexity and cost while maintaining the same effectiveness.

By systematically evaluating model performance under different contextual conditions, our aim is to determine the impact of code completeness, documentation quality, and structured contextual enhancement on vulnerability repair success rates.

Compilation errors were selectively addressed to preserve the validity of the experimental setup. Specifically, we only intervened in cases where the issue was related to a Java version mismatch. The Vul4J dataset specifies the required Java version for each project, but in some cases certain dependencies or parts of the code base relied on an older Java version, causing unexpected compatibility issues. Since predicting such discrepancies in advance was impractical, we manually resolved them by fixing these kinds of compilation errors where necessary. To minimize such issues, our prompts explicitly specified the required Java version for generated fixes. Although some cases deviated from this rule, our observations indicate that these exceptions had negligible impact on the overall repair success rate. The results of our prompt engineering process, that is, the used prompts, are published in the online appendix.

*3.3.1 CWE/CVE information.* To enhance the model's contextual understanding of vulnerabilities, we experimented with integrating CWE and CVE descriptions into the prompts where applicable. We acknowledge that this information might not be available in all practical scenarios; however, similar information can be extracted from security-related issue reports or provided by security experts.

- **CWE Descriptions** were retrieved from official sources but exhibited a high degree of variability, some were excessively detailed, while others were too concise. To ensure uniformity, we used GPT to reformat all CWE descriptions into a standardized paragraph, preserving key details while enabling potential enhancements with additional relevant information.
- **CVE Descriptions** were used as-is from the official CVE database. Then, in a later phase, we conducted an additional experiment in which the GPT was tasked with generating structured repair guidelines based on the CVE description

---
[2]https://doi.org/10.5281/zenodo.15036666

and the affected code snippet (similar to humans when reporting security-related issues).

*3.3.2 Manually derived information.* To determine which code-related contextual elements had the greatest impact on successful repairs, we manually analyzed the *vulnerable-fixed* pairs from the dataset. Our goal was to identify specific information types that could improve model performance and potentially be automatically retrieved in future retrieval-augmented repair pipelines. Among the 42 vulnerabilities, we found that 21 cases (50%) could potentially benefit from retrieving additional code or code-related information from the same project, referencing relevant library documentation, or generating detailed inline comments. For a successful repair, 6 cases (14%) required precise issue descriptions to be provided in the prompt, as the underlying vulnerabilities were highly project specific and lacked a sufficient general context. In our experiment, we used generated guidelines (by using the CVE description and vulnerable code). And finally, for 15 cases (36%) we were unable to identify additional sources of information that could be systematically retrieved to enhance the success of the repair. Examples of manual additional information (e.g., providing related features from the project or additional method signatures of the analyzed class) can be found in the online Appendix.

## 4 Results

Table 1 summarizes our findings. We systematically evaluated the efficacy of more detailed versus concise vulnerability repair prompts across multiple configurations, additionally assessing whether automatically generated documentation improves repair performance.

| Context | Run-1 | Run-2 | Run-3 | At least once |
| --- | --- | --- | --- | --- |
| GPT-4 baseline | 14 | 15 | 13 | 17 |
| GPT-4o baseline | 11 | 12 | 14 | 19 |
| Long | 13 | 13 | 14 | 17 |
| Long + Docs | 11 | 15 | 11 | 17 |
| **Short** | **13** | **13** | **13** | **18** |
| Short + Docs | 11 | 9 | 11 | 14 |
| Short + CWE | 11 | 12 | 14 | 17 |
| **Short + CVE** | **16** | **16** | **15** | **19** |
| **Short + CVE Hints + Manual** | **18** | **14** | **17** | **20** |
| Short + CVE Hints | 12 | 13 | 15 | 15 |
| Short + Manual | 9 | 13 | 10 | 13 |
| Top-3 At least once | 23 | 20 | 23 | 26 |
| GPT-4o At least once | 27 | 27 | 27 | 30 |

Table 1: Prompt results in each run and the number of distinct vulnerabilities fixed at least once

### 4.1 Long prompts

Our initial experiments used a longer system prompt with detailed vulnerability repair guidelines. We requested step-by-step explanations alongside the generated code to encourage Chain-of-Thought [18, 22] reasoning, though these were not manually inspected. As CoT is considered a useful technique in general, our prompts always requested step-by-step reasoning.

The longer prompt without inline comments (*Long*) resulted in 13-13-14 successful repairs across three runs respectively, with 5-10-8 of all the generated patches causing compilation errors (after resolving Java version inconsistencies manually). The numbers for the prompt with inline comments (*Long + Docs*) were slightly lower, with 11-15-11 successful repairs and 4-2-5 build failures. Interestingly, the introduction of inline comments correlated with slightly weaker repair performance, suggesting that additional documentation, at least in this form, did not consistently assist the model in generating correct fixes.

### 4.2 Short prompts

Next, we tested a shorter prompt, which aimed to retain key instructions while minimizing token usage for cost efficiency. The prompt without inline comments (*Short*) showed 13-13-13 successful repairs, with 9-9-8 compilation failures. The addition of inline comments (*Short + Docs*) again decreased the success rates, resulting in 11-9-11 successful repairs, with 8-7-8 compilation failures.

Our findings with **Short** and **Long** prompts indicate that the length and verbosity of the prompt did not have a significant impact on the model's ability to generate correct fixes. The overall repair success rate remained similar regardless of whether the repair logic was framed within a detailed system prompt or a compact one.

Across all experiments, most compilation errors resulted from a common set of issues:

- **Usage of non-existent methods** – The model added function calls that did not exist on the target object.
- **Unhandled exceptions** – The generated code sometimes introduced parts that required additional exception handling; however, the model did not implement the necessary code changes.
- **JVM crashes** – A few cases failed due to JVM crashes.

Although the longer prompt variants generated slightly less build errors, their repair performance is very close to the shorter prompt repair rates. Among multiple generations, the **Short** prompt managed to fix more vulnerabilities at least once than the **Long** prompt. Based on this, we decided to continue our experiments with shorter prompts only. We explored two additional variations of the shorter prompt:

- **Short + CWE**: Includes the uniform CWE description, ensuring a structured and standardized explanation. This resulted in 11-12-14 successful repairs, with 6-8-6 compilation failures.
- **Short + CVE**: Includes the raw CVE description as published in the official database. This resulted in 16-16-15 successful repairs, with 6-6-9 compilation failures.

Of the 42 vulnerability cases, 7 had no assigned CWE IDs, while only one case lacked a CVE entry. These results clearly indicate that providing CVE information significantly outperforms CWE descriptions in improving repair success. A plausible explanation is that CVE reports describe vulnerabilities in their specific real-world code, with concrete technical details. We also generated natural language repair guidelines using GPT-4o, constructed from the CVE description and the provided code snippet. These guidelines aimed to clearly summarize the essence of vulnerability and provide general best practices for mitigation. This approach also represents a more real-world setting, as the exact vulnerability description is not always available, however, various tools or human experts may generate guidelines based on the code and other available information.

The results for the **Short + CVE Hints** prompt were 12-13-15 successful repairs, with 4-6-3 compilation failures. Interestingly, despite containing the same semantic information as the raw CVE reports, this version performed worse than using raw CVE descriptions. This suggests that the model may utilize its prior knowledge more effectively when processing the CVE data in its original form.

To test whether structured manual curation could further improve repairs, we added manually collected repository and issue-related information alongside CVE-based repair guidelines **Short + CVE Hints + Manual**. The results showed a substantial improvement: 18-14-17 successful repairs, with only 3-3-4 compilation failures. This marks a notable increase compared to previous configurations, confirming that incorporating project-specific information significantly enhances the accuracy of vulnerability repair.

Although not all projects have manually extracted hint information, we ran an experiment in which only manually extracted information was provided, excluding CVE-related data **Short + Manual**. Surprisingly, this led to worse performance compared to providing no additional information: 9-13-10 successful repairs. This result suggests that, while certain project-level details are valuable, they are not sufficient on their own to guide LLM-based repair effectively. Instead, they work best in combination with external domain knowledge, such as CVE reports.

> **Answer to $RQ_1$**: Prompts without any additional information achieved <32% repair rates. By adding generated documentation, the rates reduced to <30%. CWE descriptions yielded 29.4% and CVE descriptions 37.3% success rate on average. Though GPT generated guidelines and manually extracted hints showed minimal improvement individually, their combination reached a repair rate of 38.9%.

Although no prompt could consistently outperform others, we examined aggregating high-performing prompts. Using our Top-3 prompts (leading to nine total executions across all vulnerabilities), we achieved 26 (61.9%) successful repairs in at least one run, exceeding both the original baseline and its GPT-4o reproduction (17 (40.4%) and 19 (45.2%) respectively). It is noteworthy that the baseline results generated code only three times using the same prompt. Across all 10 prompts (baseline + 9 new) with three generations each, the model successfully resolved 30 (71.4%) vulnerabilities at least once - not significantly better than the top three prompts alone. This minimal improvement might not justify the additional costs. The results emphasize that repair success varies between different prompts and that an ensemble approach, where multiple patch candidates are generated using diverse prompt configurations, might lead to significantly better result.

> **Answer to $RQ_2$**: Our Top-3 best performing prompts have successfully repaired 26 (61.9%) vulnerabilities, which is significantly higher than using individual prompts.

### 4.3 Patterns that GPT-4o Could Not Fix

During the manual context collection phase, we identified common patterns used to fix vulnerabilities. However, the results of the evaluation indicated that GPT-4o encountered difficulties in correctly applying patches to some of these patterns. The model demonstrated a limited ability to address vulnerabilities that arise from incorrect variable data types. Despite efforts to create targeted prompts to guide the model toward resolving these issues, it consistently failed to produce an acceptable solution. The model was only able to generate a correct fix when provided with explicit instructions detailing the necessary corrections. We found that forbidding the model to use certain approaches was also ineffective.

For example, in the case of VUL4J-6, an infinite loop is caused by the incorrect type of the *i* loop variable. Despite the explicit guidance, the model was unable to identify the most effective solution (changing the variable type to *long*). Although the model realized the infinite loop problem, the applied fixes, such as limiting iteration counts or capping variable values, were not sufficient. Another trivial example is VUL4J-18, where the necessary fix is explicitly stated in the documentation comment provided. However, instead of utilizing this information, the model generated its own solution. The results for the cases presented here can be found in the online appendix. This limitation questions the model's ability to fix more complex, previously unseen vulnerabilities if such a trivial problem exceeds the model's capabilities.

The model also struggles when the vulnerable implementation does not support the usually applicable patch pattern (e.g., CWE-611). We found that, in some cases, the model is prone to hallucination and uses methods that are not available for a specific object.

## 5 Threats to Validity

Our study focuses on Java, a statically typed language, which limits generalizability as repair challenges may differ in dynamic languages like JavaScript. Future work should explore such languages. LLMs' non-deterministic nature, even at low temperature settings (0.01), can cause output variations affecting results, with some runs outperforming others. This is especially noticeable in some cases, where one out of the three runs performed somewhat better than the other two.

Though we minimized manual evaluation, cases requiring human intervention might have introduced subjectivity. There may have been additional cases where minor, non-vulnerability related adjustments could have resulted in a successful fix, but that would have required a thorough inspection of each fix candidate, which is out of the scope of this study. We modified some tests to be more permissive. In each case, we were careful to make only simple modifications that did not change the semantics of the program: we caught more general exceptions instead of overly specific ones, and made allowances in examining the text messages of caught exceptions.

The extraction of additional context from the vulnerable projects was performed manually, introducing a degree of subjectivity into the process. However, we made a concerted effort to be thorough and consistent.

When comparing GPT-4 and GPT-4o, the evaluation methodology was not exactly the same. Although all results of GPT-4 [17]

were thoroughly checked manually (enhancing the code when reasonable), our study primarily relied on unit tests for evaluating GPT-4o outputs, with limited manual checks. This leaves the possibility of false negative results in the case of GPT-4o.

Finally, we chose to experiment with additional prompt contexts with GPT-4o even though the GPT-4 baseline results were better. However, the number of distinct vulnerabilities GPT-4o could fix was higher, showing a higher potential with specific prompts. Nonetheless, we plan to replicate our study with GPT-4 as well.

## 6 Conclusion

In our work, we compared the vulnerability repair performance of the GPT-4o with our previous results, using the same prompts. We found that GPT-4o performed worse than GPT-4 on average under similar conditions. However, it was able to fix a larger number of distinct vulnerabilities at least once. Therefore, to find out what additional information can help the model, we examined 9 prompts with different levels of contextual information. We generated vulnerability patches with each prompt three times for 42 vulnerable projects from the VUL4J dataset, resulting in 1,134 generated fix candidates. From our 9 zero-shot prompts, GPT-4o worked most effectively when combining generated CVE guidelines with manually extracted repository context. Based on our experiments, instead of using a single general-purpose prompt for all cases, it is more effective to use multiple different prompts to generate candidate patches, which can significantly improve the vulnerability fixing capabilities of LLMs.

In the future, we would like to use a RAG-based solution to get additional information, as well as develop an artificial intelligence-based system that dynamically creates the prompts based on the vulnerability and project information.

## Acknowledgments

This work was supported in part by the European Union project RRF-2.3.1-21-2022-00004 within the framework of the Artificial Intelligence National Laboratory; and in part by the Project no TKP2021-NVA-09 implemented with the support provided by the Ministry of Culture and Innovation of Hungary from the National Research, Development and Innovation Fund, financed under the TKP2021-NVA funding scheme. The work has also received support from the European Union Horizon Program under the grant number 101120393 (Sec4AI4Sec).